\documentclass[journal]{new-aiaa}
\usepackage[utf8]{inputenc}
\usepackage{textcomp}

\usepackage{graphicx}
\usepackage{amsmath}
\usepackage[version=4]{mhchem}
\usepackage{longtable,tabularx}
\usepackage{bm}

\usepackage{background}
\backgroundsetup{
  position=current page.east,
  angle=-90,
  nodeanchor=east,
  vshift=-4mm,
  hshift=30mm,
  opacity=1,
  scale=3,
  contents=\textsc{PREPRINT} -- https://doi.org/10.2514/1.B38346
}

\title{Numerical investigation of reversed gas feed configurations for Hall thrusters}

\author{Stefano Boccelli \footnote{PhD Candidate, Department of Aerospace Science and Technology, via La Masa 34, 20156 Milano.}}
\affil{Polytechnic University of Milan, Milan, Italy 20156}
\author{Thierry E. Magin \footnote{Professor, Aeronautics and Aerospace Department, Waterloosesteenweg 72, B-1640 Sint-Genesius-Rode, Senior AIAA member.}}
\affil{von Karman Institute for Fluid Dynamics, Sint-Genesius-Rode, Belgium 1640}
\author{Aldo Frezzotti \footnote{Full Professor, Department of Aerospace Science and Technology, via La Masa 34, 20156 Milano.}}
\affil{Polytechnic University of Milan, Milan, Italy 20156}

\begin{document}

\maketitle

\begin{abstract}
A reversed gas feed configuration for Hall thrusters is proposed and studied numerically. 
As an alternative to standard direct injection, we investigate the effect of injecting the propellant near the channel exit and directing it backwards, towards the anode.
The resulting neutral density and average velocity fields are studied with the Direct Simulation Monte Carlo method for both cold and warm anode conditions.
The residence time of neutral particles inside the channel and in the ionization region is computed using a test-particle Monte Carlo method.
The computations indicate that the reversed injection allows for increasing the mass utilization efficiency of a standard feed configuration from 2--5\% to a maximum of 20--30\% depending on the initial efficiency.
\end{abstract}

\section{Introduction}
\lettrine{H}{all} effect thrusters (HET), also known as stationary plasma thrusters (SPT) \cite{zhurin_physics_1999,morozov2000fundamentals}, are effective devices for attitude control and maneuvering of spacecraft, and are envisaged as propulsion systems for future manned missions inside the solar system.
The benefit of employing electric propulsion engines, as opposed to chemical thrusters, lies in the high specific impulse, that exceeds 1000~s and can reach 5000~s or more for high specific impulse HETs \cite{goebel_fundamentals_2008,hofer2006high}.
Among the figures of merit, the mass utilization efficiency (or propellant efficiency) $\eta_m$ plays a crucial role \cite{brown2009methodology}:
\begin{equation}
    \eta_m = \dot{m}_i \, /\dot{m}_p 
\end{equation}

\noindent with $\dot{m}_i$ the mass flux of ions ejected by the thruster and $\dot{m}_p$ the injected flow of propellant.
If a hollow cathode is used to neutralize the plume, then $\dot{m}_p$ includes an additional cathode loss contribution.
In this work, we neglect this for simplicity, assuming that no propellant is lost through the cathode.
This would actually be the case if a filament cathode is employed, as frequently done in experiments \cite{guerrini1997parameter}.

The mass utilization efficiency varies widely with the thruster size and operating conditions \cite{raitses1998propellant,komurasaki1996performance,kurzyna2019preliminary}, and can range from below 0.5 for micro-thrusters up to above 0.9.
The gas injector design plays a crucial role in the mass utilization efficiency, as it influences the residence time of neutral particles inside the chamber, and thus the possibility that neutrals are ionized by electron impact collisions.
A large number of gas feed configurations have been tested throughout the years \cite{reid2007review}, including injection ports located at the middle of the channel \cite{garrigues2004optimized} and various anode configurations.


With possible exceptions inside the gas injector, the neutral gas in the Hall thruster channel is characterized elsewhere by strong degrees of rarefaction, as expressed by the Knudsen number $\mathrm{Kn} = \lambda/L_{\mathrm{tc}}$, where $\lambda$ is the particle mean free path and $L_{\mathrm{tc}}$ the characteristic size of the thruster channel.
Besides the channel size, the value of $\mathrm{Kn}$ for Hall thrusters also strongly depends on the mass flow rate.
In this work we consider the SPT-100 thruster \cite{mitrofanova2011new}, where $100$ denotes the outer diameter in mm and with an inner diameter of 70 mm.
For such thruster, the Knudsen number based on the channel width lies roughly in the range of 0.1 to 1, as will be discussed in Appendix~\ref{sec:appendix-Knudsen-num}.
The neutral flow is therefore transitional and shall be analyzed by use of the kinetic theory of gases \cite{koganrarefied}, employing for example the Direct Simulation Monte Carlo (DSMC) method \cite{bird1994molecular}.
Such method can treat equally accurately both collisional and collisionless regimes.
Other collisionless approaches such as view-factor methods (see for example \cite{araki2020radiosity}) could prove useful for low mass flow rates or smaller devices such as the SPT-50 or SPT-20 thrusters \cite{guerrini1997parameter}, characterized by a larger surface-to-volume ratio.

\subsubsection*{Aim and structure of this work}

In this work, we investigate via numerical simulation the possibility to employ a reversed gas feed configuration: 
the propellant is injected near the outer end of the channel and directed inwards, towards the anode (see Fig.~\ref{fig:injection-simplified-model}).
Such strategy clearly presents a number of additional technical challenges in the design of the gas feed system, if compared to a direct injection.
The present study does not consider such technical difficulties, but merely investigates the resulting neutral gas dynamics, once injected into the chamber.
For simplicity, all plasma effects will be neglected, except for the increased temperature of the anode \cite{book2010effect,martinez2013propellant}.
Electron-neutral and ion-neutral collisions will be neglected, although these are known to affect the neutral temperature and momentum.
Despite the heavy simplifications, results will provide useful insights on the neutral behavior for the considered conditions.

This work is structured as follows.
At first, a heavily simplified analysis of the problem is discussed in Section~\ref{sec:reversed-inj-simplified}, and the ionization probability is written in terms of the 
neutral residence time in the ionization region.
In Section~\ref{sec:DSMC-simulations} the neutral flow field is investigated more in detail for direct and reversed injection configurations using DSMC simulations.
In Section~\ref{sec:single-particle-monte-carlo} a test-particle Monte Carlo algorithm is implemented and parallelized on a GPU.
The algorithm follows single particles in the thruster channel, moving in a pre-computed DSMC solution (for collisional conditions) or in a vacuum (for collisionless cases).
Such algorithm is employed to study the average residence time of particles in different configurations in Section~\ref{sec:single-particle-results}.
Distributions and average values of residence times are discussed.
Finally, the obtained residence times are employed to study the increased mass utilization efficiency for the reversed injection against a direct injection efficiency.
Conclusions are drawn in Section~\ref{sec:conclusions}.


\section{Direct vs reversed injection: simplified analysis}\label{sec:reversed-inj-simplified}

\begin{figure}[htpb]
    \centering
    \includegraphics[width=0.6\columnwidth]{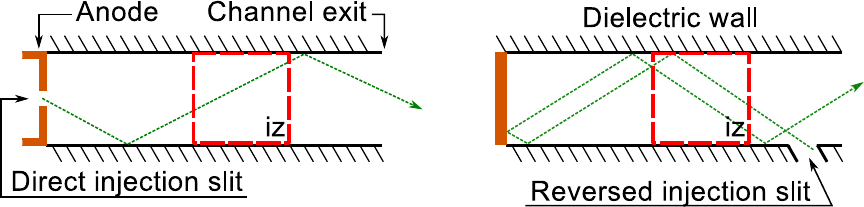}
    \caption{Single particle trajectory with specular wall scattering.
    Left: direct injection from the anode.
    Right: reversed injection.
    Ionization region is labeled ``iz''.}
    \label{fig:injection-simplified-model}
\end{figure}

A strongly simplified view of the effect of reversed injection can be obtained via a preliminary model, where we assume (i) monoenergetic particles (no thermal velocity), (ii) free molecular conditions and (iii) specular wall collisions. 
More representative conditions will be addressed in the following sections.
The problem is depicted in Fig.~\ref{fig:injection-simplified-model}.
Assuming monoenergetic neutrals implies that all particles have the same initial axial velocity $\varv_z$, and therefore all neutrals injected inwards will enter the channel.
In free molecular flow conditions, gas phase collisions are negligible:
every neutral moves independently from the others, and its residence time inside the thruster channel depends on its velocity and on the interaction with the walls.
If we additionally assume a specular reflection wall-collision model (zero energy and momentum accommodation), the time of permanence inside the ionization region for neutral atoms injected from the anode $\tau^{\mathrm{dir}}$ (Fig.~\ref{fig:injection-simplified-model}-(a)) simply reads $\tau^{\mathrm{dir}} = L_{iz} \, / \varv_z$, where $L_{iz}$ is the length of the ionization region.
On the other hand, by injecting a neutral from the open end of the channel (with the same value of the axial velocity, but opposite sign) such particle would cross the ionization region twice, and the residence time would double: $\tau^{\mathrm{rev}} = 2 \, \tau^{\mathrm{dir}}$.
Since electron-neutral ionizing collisions can be assumed to be uncorrelated events, the probability $P_{iz}$ that a given neutral particle ionizes during the time interval $\tau$ is
\begin{equation}\label{eq:Piz-1-exp-nu-tau}
    P_{\mathrm{iz}} = 1 - \exp\left[- \nu_{\mathrm{iz}} \tau\right] \, ,
\end{equation}

\noindent where $\nu_{\mathrm{iz}}$ is the electron-neutral ionization frequency, that depends on the local electron density, temperature and in general on their distribution function \cite{boccelli2020electrons}.
For the goals of this work, it is not necessary to specify $\nu_{\mathrm{iz}}$ further.
The ionization probability in the reversed injection case can thus be obtained from the direct injection case as
\begin{equation}\label{eq:Piz-rev-direct-simple}
    P_{\mathrm{iz}}^{\mathrm{rev}} = 1 - \left(1 -  P_{\mathrm{iz}}^{\mathrm{dir}} \right)^{\tau^{\mathrm{rev}}/\tau^{\mathrm{dir}}} \, .
\end{equation}

In this simplified model, we have $\tau^{\mathrm{rev}}=2\tau^{\mathrm{dir}}$, such that $P_{\mathrm{iz}}^{\mathrm{rev}}\ge\,P_{\mathrm{iz}}^{\mathrm{dir}}$ for all conditions, with the largest improvements for smaller ionization probabilities.

Clearly, this model is heavily simplified.
In a more realistic scenario, the non-zero temperature of neutrals is such that the residence times are characterized by a distribution that depends on the geometry.
Each different particle has a different ionization probability, slow particles being naturally more likely to ionize since they spend more time inside the ionization region.
Moreover, due to the finite temperature, a fraction of particles injected near the channel end will be lost in the plume prematurely.
Such particles will constitute in a loss in terms of residence time.

The effect of wall collisions also strongly affects the dynamics of the problem, according to the value of the accommodation coefficient \cite{koganrarefied}.
This may significantly depart from 0 (specular reflection) and approach unity (diffuse collisions), limiting the mobility along the channel.
Finally, the presence of gas-phase collisions would have the effect of pushing atoms towards the exit.
The relative importance of these phenomena depends on the Knudsen number, and thus on the thruster size and mass flow rate.
All mentioned effects will be considered in the remaining of this work for two direct injection cases and for various reversed injection configurations.


\section{DSMC simulations}\label{sec:DSMC-simulations}

In this section, we perform DSMC simulations of direct and reversed injection strategies, as to obtain density, velocity and temperature fields of the neutral gas.
These simulations are complementary to the simplified analysis of Section~\ref{sec:reversed-inj-simplified}, and directly include the effect of finite thermal velocity, diffuse wall collisions and gas-phase collisions.

\begin{figure}[htpb]
    \centering
    \includegraphics[width=0.5\columnwidth]{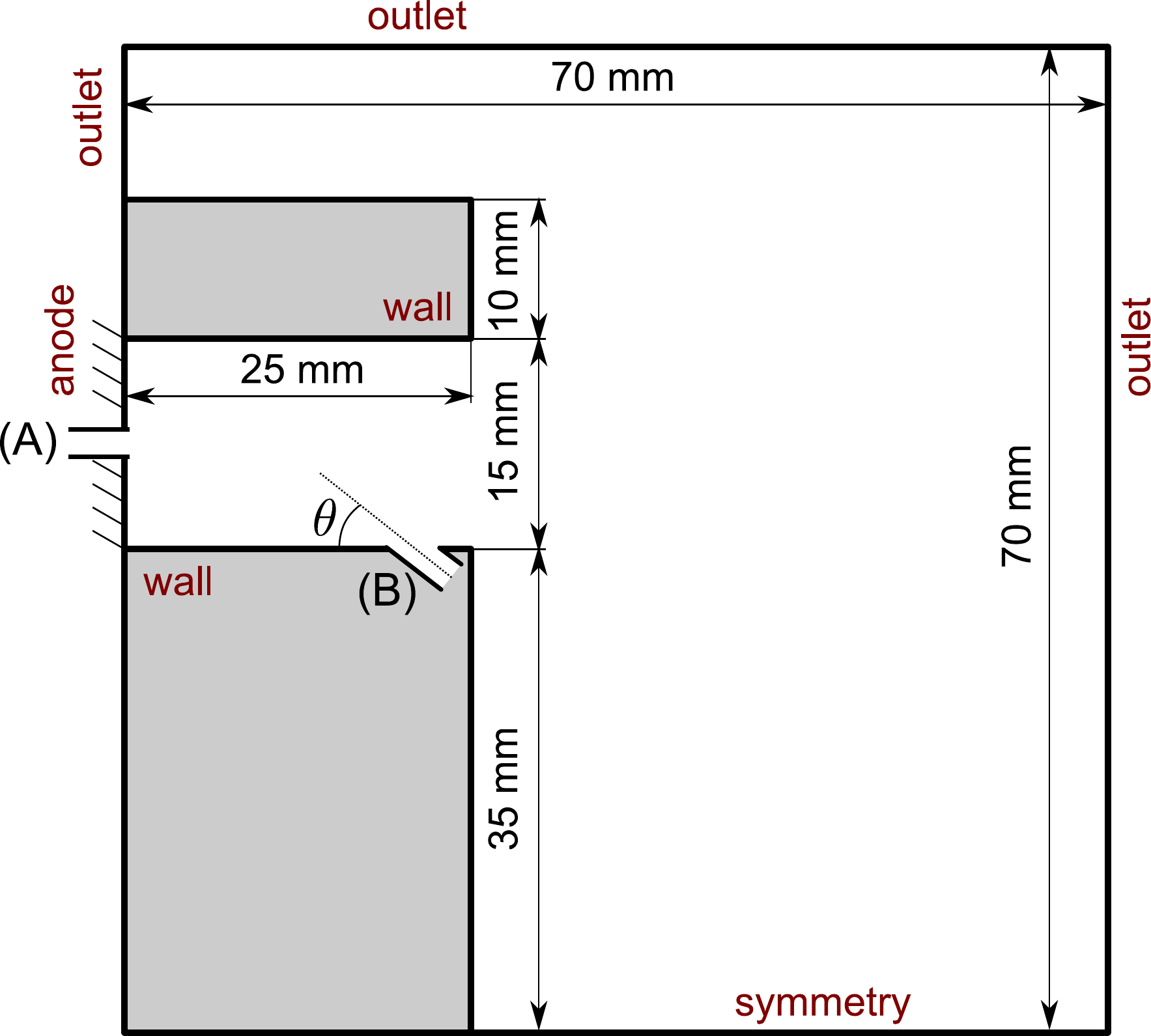}
    \caption{Computational domain and boundary conditions for the SPT-100 DSMC simulations. Axial-radial plane.}
    \label{fig:DSMC-geometry}
\end{figure}

The simulations are performed with the SPARTA DSMC code \cite{plimpton2019direct} on a 2D computational domain with axial symmetry, and represents a slice in the radial-axial plane $R$--$z$, as shown in Fig.~\ref{fig:DSMC-geometry}.
The considered geometry is characteristic of the SPT-100 thruster \cite{mitrofanova2011new}, and the propellant mass flow rate is fixed at $\dot{m}_n = 5$ mg/s, injected either uniformly from the whole anode (case A1), or from a small slit in the anode (case A2).
In cases ``B'' the reversed injection is considered, with angle $\theta$ with respect to the axial direction ($\theta=30^\circ$ is a radial injection), and a slit centered at axial position $z_s$.
Cases B30\_1, B60\_1 and B90\_1 consider different injection angles of $\theta=30^\circ,60^\circ$ and $90^\circ$ respectively, for a given slit position located close to the exit plane.
In cases B30\_2, B30\_3 and B30\_4, we keep the angle $\theta = 30^\circ$ and progressively move the slit towards the anode.

\begin{table}[hbt!]
\caption{\label{tab:reversed-injection-cases}
Test cases for the reversed injection configuration.}
\centering
\begin{tabular}{cccc}
\hline
      $z_s$ = 22.5 mm &  18.5  &  15.5  &  12.5  \\
     \hline
     B30\_1        &   B30\_2    &  B30\_3 & B30\_4 \\
     B60\_1        &   -        &  -  &  - \\ 
     B90\_1        &   -        &  -  &  - \\
     \hline
\end{tabular}
\end{table}

Xenon gas is considered.
All cases are run for two different conditions of wall temperature $T_w$ and anode temperature $T_a$:
\begin{itemize}
    \item Cold anode: $T_a = T_w = 300$ K;
    \item Warm anode: $T_a = 1000$ K, $T_w = 300$ K.
\end{itemize}

\noindent Actual wall and anode temperatures for real thrusters strongly depend on the operating conditions and require some hours of operation before reaching a steady state \cite{mazouffre2006calibrated,huang2011neutral}.
In cases A1 for cold and warm anode, atoms are injected from radial position $R = 35$ to $50$ mm, assuming a Maxwellian distribution with zero average velocity and the anode temperature.
The inlet flux is thus completely thermal, and the number density is set as to result in the required mass flux.
The slit size is 1 mm and for the B cases is located on the inner wall.
The boundary condition to reproduce a physically accurate orifice expansion requires particular care (see for example \cite{sharipov2004numerical}) and ultimately depends on the detailed structure of the gas feed.
For simplicity, we assume sonic injection, at the anode temperature in case A2 and at wall temperature for cases B.
Particles injection is therefore performed by sampling a Maxwellian distribution with sonic average velocity, equal to $177.9$ m/s for a temperature of $300$ K and $324.8$ m/s for injection temperature of $1000$ K.
For case A2 the average injection velocity is axial, while for cases B its components form an angle $\theta$ with respect to the side walls.
The number density is finally chosen as to give the required mass flux.

\begin{figure*}[htpb]
    \centering
    \includegraphics[width=\textwidth]{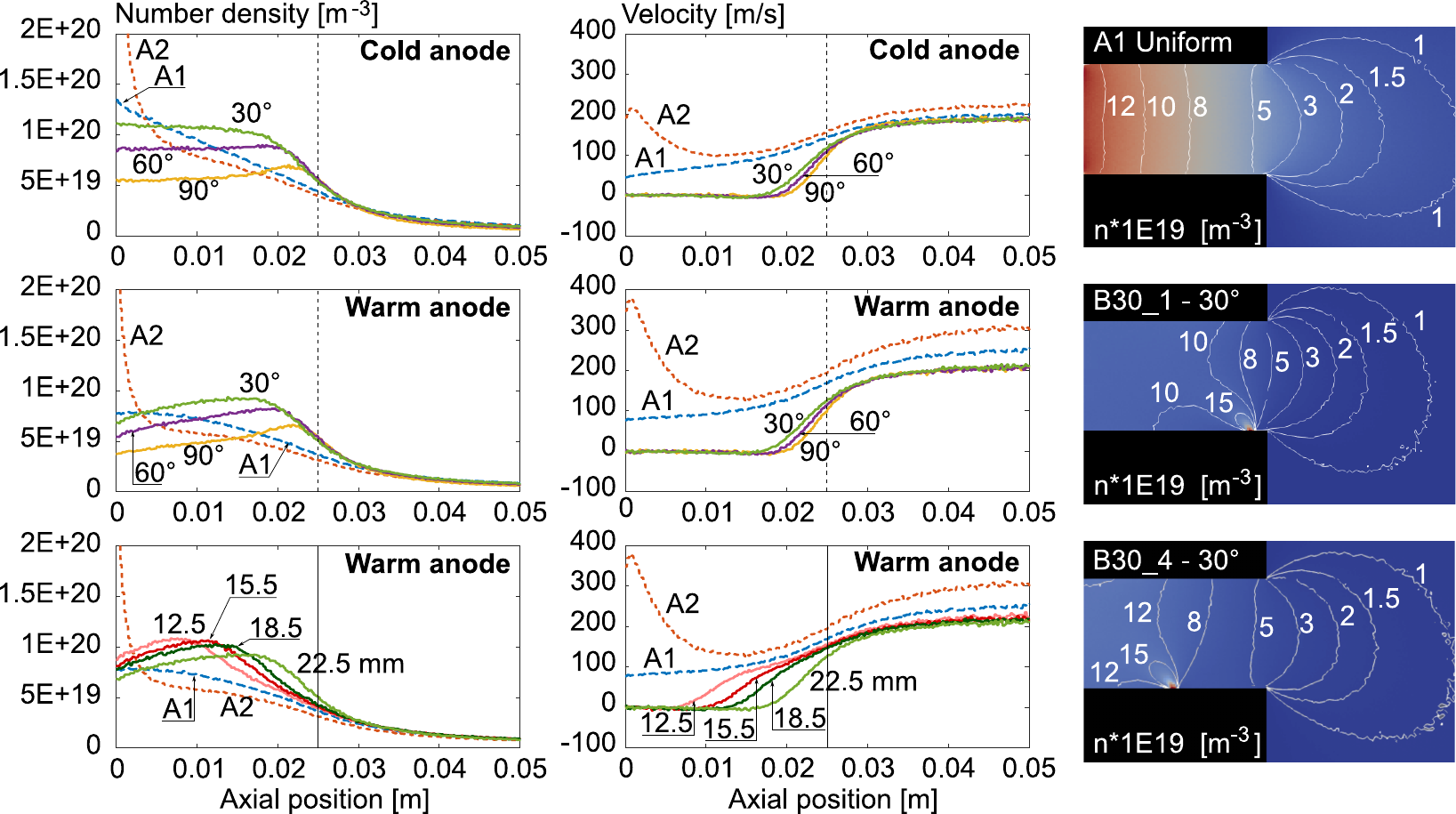}
    \caption{DSMC simulations. Left and Center: density and velocity at the channel centerline.
    A1, A2: direct injection. B: reversed injection. 
    Right column: number density contours for cases A1, B30\_1 and B30\_4, cold anode.}
    \label{fig:DSMC-results}
\end{figure*}

The DSMC cell dimension is chosen as to resolve the local mean free path, and a discretization of 280$\times$280 cells in the whole domain proved adequate.
The time step is set to 1 $\mu$s as to properly resolve the mean free time.
The number of simulated particles in steady state is in the order of $150\,000$, providing sufficient resolution inside the channel and in the first part of the plume.
A variable soft sphere (VSS) collision model is employed \cite{bird1994molecular}, where the cross section is $\sigma = \pi d^2$, and the 
diameter $d$ depends on the relative velocity $g$,
\begin{equation}
  d = d_{\mathrm{ref}} \left[ \frac{1}{\Gamma(5/2 - \omega)} \left(   \frac{ 2 k_B T_{\mathrm{ref}}}{m_r g^2}\right)^{\omega - 1/2} \right]^{1/2} \, ,
\end{equation}

\noindent with relative mass $m_r$.
The deflection angle $\chi$ reads
\begin{equation}
  \chi = 2 \cos^{-1} \left[ \left(b/d \right)^{1/\alpha} \right] \, ,
\end{equation}

\noindent where $b$ is the impact parameter.
In this work, we use a reference diameter $d_{\mathrm{ref}} = 5.65 \times 10^{-10}\ \mathrm{m}$ at the reference temperature $T_{\mathrm{ref}} = 273.15\ \mathrm{K}$, viscosity index $\omega = 0.85$ and scattering parameter $\alpha = 1.44$ (see \cite{bird1994molecular}).
Each simulation takes roughly 15 minutes on a laptop (Intel i5 processor, single core computation).
The resulting density and velocity fields are shown in Fig.~\ref{fig:DSMC-results} for the cases A1, A2, and for the angled injection cases of Table~\ref{tab:reversed-injection-cases}.
In all cases, the effect of the anode temperature is seen to decrease the density and increase the average velocity, as expected.
The reversed injection strategy appears to have a strong impact on the averaged fields.
All three injection angles level out the density inside the chamber and bring the average velocity to zero, with the injection at $30^\circ$ providing the highest density.
Referring to the center column in Fig.~\ref{fig:DSMC-results}, a low (zero in this case) average velocity can be obtained by the reversed injection approach, suggesting that the average residence time of neutrals can be increased.
In terms of flux balance, the reversed injection configuration mimics the flow in a closed end channel, where the global mass flux is zero.        
Still, individual particles reach the anode via thermal motion, eventually being reflected back and leaving the channel.
The rarefied conditions are such that the average fields only show a marginal picture of the problem, and a more in-depth analysis of residence times should be performed (see sections \ref{sec:single-particle-monte-carlo} and \ref{sec:single-particle-results}).

Observing the $\theta = 30^\circ$ density contour in the right column of Fig.~\ref{fig:DSMC-results}, it appears that the actual direction of injection is much closer to $45^\circ$.
This can be explained by considering that the velocity field is an average of (1) the population of injected particles at $30^\circ$ and (2) the population of scattered particles that are leaving the domain.
For injections at very small angles, the second effect can be expected to increase, as the fraction of particles colliding with the inner wall increases.

The effect of shifting the slit towards the anode is shown in Fig.~\ref{fig:DSMC-results}-Bottom row.
As could be expected, the density and velocity profiles approach the uniform injection case A1, as the slit position approaches the anode. 
For such cases, only the warm anode case is shown. Results for the cold anode case are qualitatively analogous and will not be shown here.

\subsection{Effect of ionization and ion-neutral interactions on the DSMC fields}

As mentioned, ionization has been neglected in this study.
For this reason, we can expect the density profiles of Fig.~\ref{fig:DSMC-results} to be somehow over-predicted, especially
after neutrals have crossed the ionization region.
The accuracy of our simulations ultimately depends on the operating conditions, in terms of magnetic field and applied potential.
For example, Koo and Boyd \cite{koo2003computational} report the neutral density profiles for an SPT-100 thruster 
in presence of ionization, for a mass flow rate comparable to this work.
In their simulations, the neutrals density appears to be strongly affected by ionization, differing by up to a factor 10 
from our neutral results, in some parts of the domain.

It should be noted that, if one considers completely collisionless operating conditions, these potentially large discrepancies would 
have no effect in terms of neutral residence times, since collisionless conditions make an accurate prediction of the baseline fields completely unnecessary.
%
In the situations considered in this work, collisionality appears not to be the predominant factor, as also discussed in Appendix~\ref{sec:appendix-Knudsen-num}.
Although it does play a significant role (as will be further discussed in Section~\ref{sec:collisional-single-particle}), we could expect our results to be 
anyway indicative of the general behavior.
In any case, more complete simulations are suggested as a future work to investigate this to a full extent.

Secondly, ion-neutral interactions (either charge exchange -- CEX -- or momentum exchange -- MEX) can be expected to modify both the 
baseline fields and the average residence time of neutals in the ionization region.
This effect becomes more important for higher ion densities and higher acceleration potentials.  
To get an estimation of the importance of such effect, we propose here a simple analysis of the Knudsen number based on the ion-neutral collisionality.
We denote by $\nu_{ni}=n_i\sigma_{ni}V_i$ the frequency at which a given neutral particle is hit by a fast-moving ion.
We consider an ion density $n_i\approx10^{18}~\mathrm{m^{-3}}$, a cross-section $\sigma_{ni}\approx5\times10^{-19}~\mathrm{m^2}$ and an ion velocity $V_i \approx 10~\mathrm{km/s}$ 
(ions are created at a low velocity and typically accelerate up to, say, $15-20~\textrm{km/s}$, see for example \cite{boccelli2020collisionless}),
giving $\tau_{ni}\equiv \nu_{ni}^{-1}\approx2\times10^{-4}~\mathrm{s}$.
The residence time of a neutral particle inside the channel can be roughly estimated as $\tau_n\approx L_{\mathrm{tc}}/V_n$, with $L_{\mathrm{tc}}=0.025~\mathrm{m}$ 
and $V_n\approx100~\mathrm{m/s}$, such that $\tau_n\approx1.5\times10^{-4}~\mathrm{s}$.
From this simplified analysis, the mean ion-neutral collision time and the neutral transit time appear comparable $\tau_{ni}\approx\tau_{n}$.
Therefore, ion-neutral collisions are not the predominant contribution to the problem, but completely neglecting them,
as we do in the present work, is a strong assumption, that should be relaxed in a future work.
 
%
%


\section{Test-particle Monte Carlo simulations }\label{sec:single-particle-monte-carlo}

Obtaining the residence time of neutral particles inside a region could be easily done by the DSMC method.
However, such feature is usually not included in standard software, that would need in most cases some ad-hoc modifications, such as attaching one additional field --namely, the residence time-- to the particle object, and updating it during the particle motion.
In the present work we pursue a different strategy, referred to as test-particle Monte Carlo (TPMC), where a single test particle is moved in a background of gas, whose properties are given (known from the previous DSMC simulations).

The procedure goes as follows. 
First, a target particle is injected from the gas feed, with velocity sampled from a Maxwellian distribution at the injection conditions.
Then, the particle is moved ballistically for a time step, detecting wall collisions, if any.
The test particle collides with the background (assumed Maxwellian) with a probability that depends on the local density and temperature fields.
Finally, when the particle leaves the channel, the residence time is saved and a new target particle is processed.

This strategy is ultimately a simplified version of the DSMC method, where particle collisions happen with a given background and not between simulated particles.
As a result, the scheme implementation is trivial and heavily parallelizable.
This procedure is particularly suited for describing free molecular flows, where one simply has to remove the gas-collision phase.
This can be useful for lower propellant flow rates and/or thrusters with a larger surface-to-volume ratio, when the Knudsen number is large (see Appendix~\ref{sec:appendix-Knudsen-num}).
No DSMC computations for the background are needed in this case.

\begin{figure}[htpb]
    \centering
    \includegraphics[width=0.25\columnwidth]{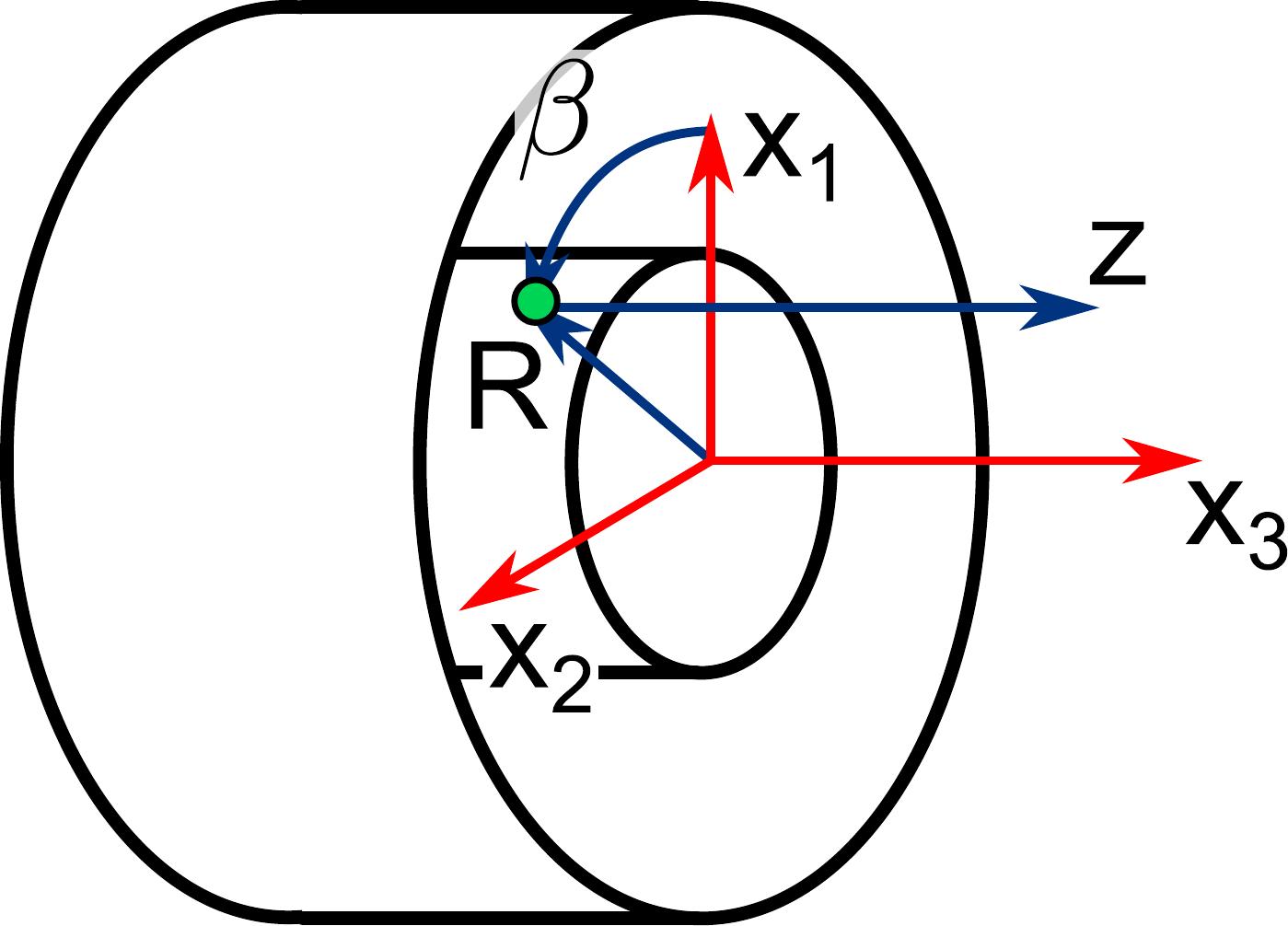}
    \caption{Cartesian and cylindrical reference systems for the test-particle Monte Carlo simulations.}
    \label{fig:single-particle-ref-sys}
\end{figure}

The reference system for the simulations is shown in Fig.~\ref{fig:single-particle-ref-sys}.
Advection is performed in the three directions $(x_1, x_2, x_3)$, and after each time step, the particle position is rotated back into the $(x_1,x_3)$ plane, as to impose axial symmetry.
First, cylindrical coordinates are found by
\begin{subequations}
\begin{align}
    R &= \left[ x_1^2 + x_2^2\right]^{1/2} \ , \\
    \beta &= \mathrm{atan}(x_2/x_1) \ , \\
    z &\equiv x_3 \ .
\end{align}
\end{subequations}

The particle position vector is rotated back to the symmetry plane by the angle $\beta$,
\begin{equation}
    \begin{pmatrix}
       x_1^\prime \\ x_2^\prime \\ x_3^\prime
    \end{pmatrix}
    =
    \begin{pmatrix}
       R \\ 0 \\ x_3
    \end{pmatrix} 
\end{equation}

\noindent and the same rotation is performed on the particle velocity vector $\bm{\varv}$:

\begin{equation}
    \begin{pmatrix}
       \varv_1^\prime \\ \varv_2^\prime \\ \varv_3^\prime
    \end{pmatrix}
    =
    \begin{bmatrix}
    \cos{\beta} & \sin{\beta}  & 0 \\
    -\sin{\beta} & \cos{\beta} & 0 \\
    0  &  0  &  1 \\
    \end{bmatrix}
    \begin{pmatrix}
       \varv_1 \\ \varv_2 \\ \varv_3
    \end{pmatrix} \, .
\end{equation}

When a particle crosses the walls during the advection phase, a wall-collision is detected:
the particle is reflected into the domain, sampling its new velocity from a half-Maxwellian, characterized by the wall temperature.
This coincides with assuming a diffusive scattering kernel, with unitary accommodation coefficient.
The new particle position after the wall scattering event is computed accordingly.

The algorithm is finished if one wishes to consider a collisionless gas and only particle-surface interactions. In the collisionless case, the resulting residence times would scale linearly with the thruster size, and results obtained for a given configuration could easily be extrapolated to scaled geometries.
Residence times can also be expected to scale with the square root of the surface temperatures, with higher temperatures causing higher thermal velocities and lower residence times.
However, having a different anode and wall temperatures make the scaling less trivial.

\begin{figure}[htpb]
    \centering
    \includegraphics[width=0.7\columnwidth]{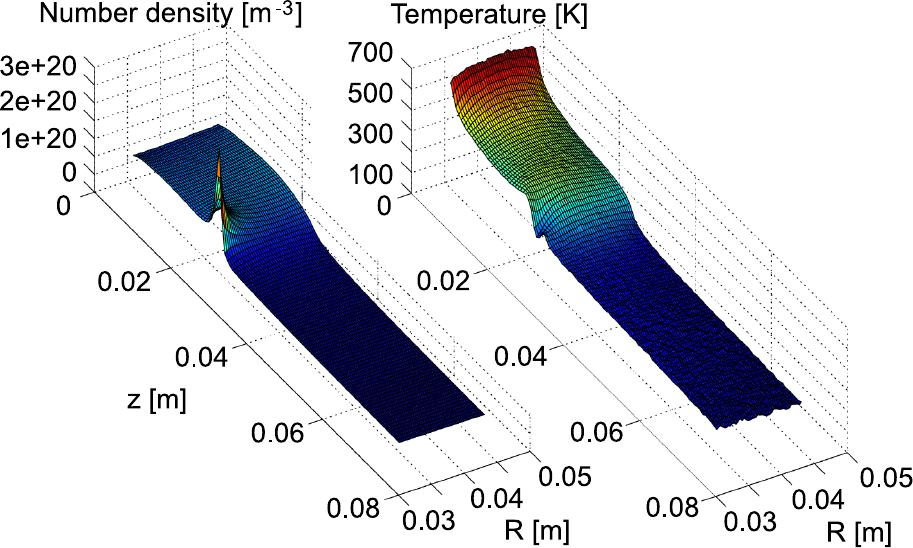}
    \caption{Density and temperature profiles for case B30\_1, warm anode, extracted from a strip of the DSMC simulation.}
    \label{fig:single-particle-field-plots}
\end{figure}

Collisions of the target gas with the background are performed as follows.
First, density, velocity and temperature are imported from the DSMC simulation (see for example some fields in Fig.~\ref{fig:single-particle-field-plots}).
The values are provided as tabulated on the $z-R$ plane as a 2D matrix and are sampled at the current particle location via 
bilinear interpolation (see for example \cite{william2007numerical}).
This gives the local background gas density $n_{BG}$, average velocity $\vec{\varv}_{BG}$ and temperature $T_{BG}$.
The relative velocity between the particle and the background is $g = |\vec{\varv} - \vec{\varv}_{BG}|$.

%
%

%
%
%

Collisions are then evaluated assuming a hard-sphere (HS) collision model.
For a Maxwellian background and HS collisions, the velocity-dependent collision frequency is easily derived \cite{cercignani1988boltzmann} and reads
\begin{equation}
  \nu(g) = \frac{\sigma n_{BG}}{\pi} \sqrt{\frac{2 \pi k_B T_{BG}}{m}} \, \psi\left(g \sqrt{\frac{m}{2 k_B T_{BG}}} \right) \, ,
\end{equation}

\noindent with $\sigma = 1.0351\times10^{-18} \ \mathrm{m^2}$ the xenon hard-sphere cross-section from \cite{bird1994molecular}, $k_B$ the Boltzmann constant and $m$ the mass of xenon background particles.
The function $\psi$ reads
\begin{equation}
   \psi(x) \equiv e^{-x^2}  + \left( 2 x + \frac{1}{x} \right) \frac{\sqrt{\pi}}{2} \mathrm{erf}(x) \, .
\end{equation}

\noindent For consistency with the DSMC computations, one could employ the variable-hard-sphere (VHS) or the variable-soft-sphere (VSS) collision models instead of the simpler HS.
In such cases, expressions of the velocity-dependent collision frequency are not trivial and one should instead compute it numerically.
For the considered conditions, we have verified by numerical simulation that employing such models does not change considerably our results.
Therefore, we only consider the HS model here.
From the collision frequency, the probability that the test particle experiences a collision during the time step $\Delta t$ is then computed by (see for example \cite{vahedi1995monte})
\begin{equation}
    P_{c} = 1 - \exp\left( - \nu \Delta t \right) \, .
\end{equation}

It should be stressed that the time step needs to ensure a collision probability $P_c$ reasonably smaller than 1.
Since the collision frequency is position-dependent, one possibility consists in updating the time step along the particle trajectory.
A collision is performed if a random number $\mathcal{R}$ satisfies the condition: $\mathcal{R} < P_c$.
If it is the case, then a neutral colliding partner is sampled from the background Maxwellian distribution and a collision is performed
by rotating the relative velocity vector by a random angle \cite{garcia2000numerical}.

The discussed algorithm was implemented in two versions, one serial and one exploiting GPU acceleration with CUDA, available at \cite{boccelli2020}.
As all particles are independent, the algorithm is heavily parallelizable.
Running the CUDA version on an NVidia GTX 760 GPU, it was possible to obtain speedups of over 1600 times, for $10^6$ simulated particles, with respect to the serial program, running on an Intel CORE i5 processor with 12 GB of RAM.
The wall time for simulating 1 million particles on a background field of $50 \times 100$ grid points is in the order of a few seconds for the parallel code.


\section{Results of test-particle Monte Carlo simulations}\label{sec:single-particle-results}

\begin{figure}[htpb]
    \centering
    \includegraphics[width=0.5\columnwidth]{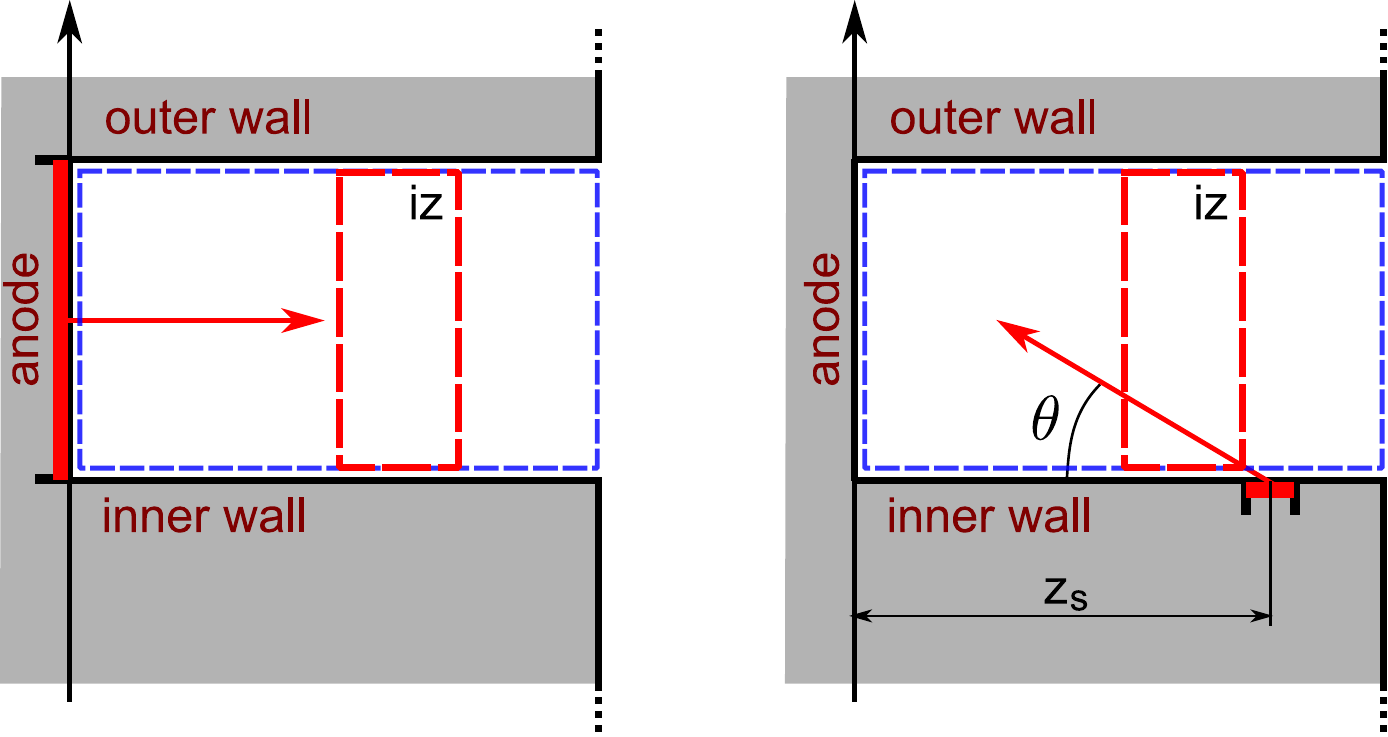}
    \caption{Computational domain for the test-particle Monte Carlo simulations (blue dashed box) and ionization region ``iz'' (red dashed box). Left: direct injection. Right: reversed injection.}
    \label{fig:single-particle-monte-carlo-domain}
\end{figure}

The scheme sketched in Section~\ref{sec:single-particle-monte-carlo} is here employed to analyze the residence time of neutrals.
The computational domain only includes the SPT-100 channel, such that particles that reach the plume cannot be scattered back and are effectively lost.
Inside the channel, ionization typically happens in a region located before the exit plane.
For the considered geometry, we assume the ionization region is located between $x = 0.013$ and $0.018 \ \mathrm{m}$ (see \cite{ahedo2001one}).
The computational domain and the ionization region are shown in Fig.~\ref{fig:single-particle-monte-carlo-domain}.
For a better clarity, the distribution is shown for the \textit{logarithm} of residence times $\tau$.

First, we consider in Section~\ref{sec:free-molec-single-particle} the free molecular limit, and run the test-particle Monte Carlo neglecting the gas-phase collision step.
This step allows to observe some physical features that would be difficult to notice in the collisional case.
Collisions are then introduced in Section~\ref{sec:collisional-single-particle}, providing a more reliable figure of merit and the average residence time inside the ionization region is computed.
Finally, this data is used in Section~\ref{sec:estimate-efficiency} to estimate the effect of reversed injection on the mass utilization efficiency.
In all simulations, the particles injection is performed as described for the DSMC case (Section \ref{sec:DSMC-simulations}).
Computations are performed on $10^6$ test particles.


\subsection{Free-molecular limit}\label{sec:free-molec-single-particle}

\begin{figure*}[!htpb]
    \centering
    \includegraphics[width=0.95\textwidth]{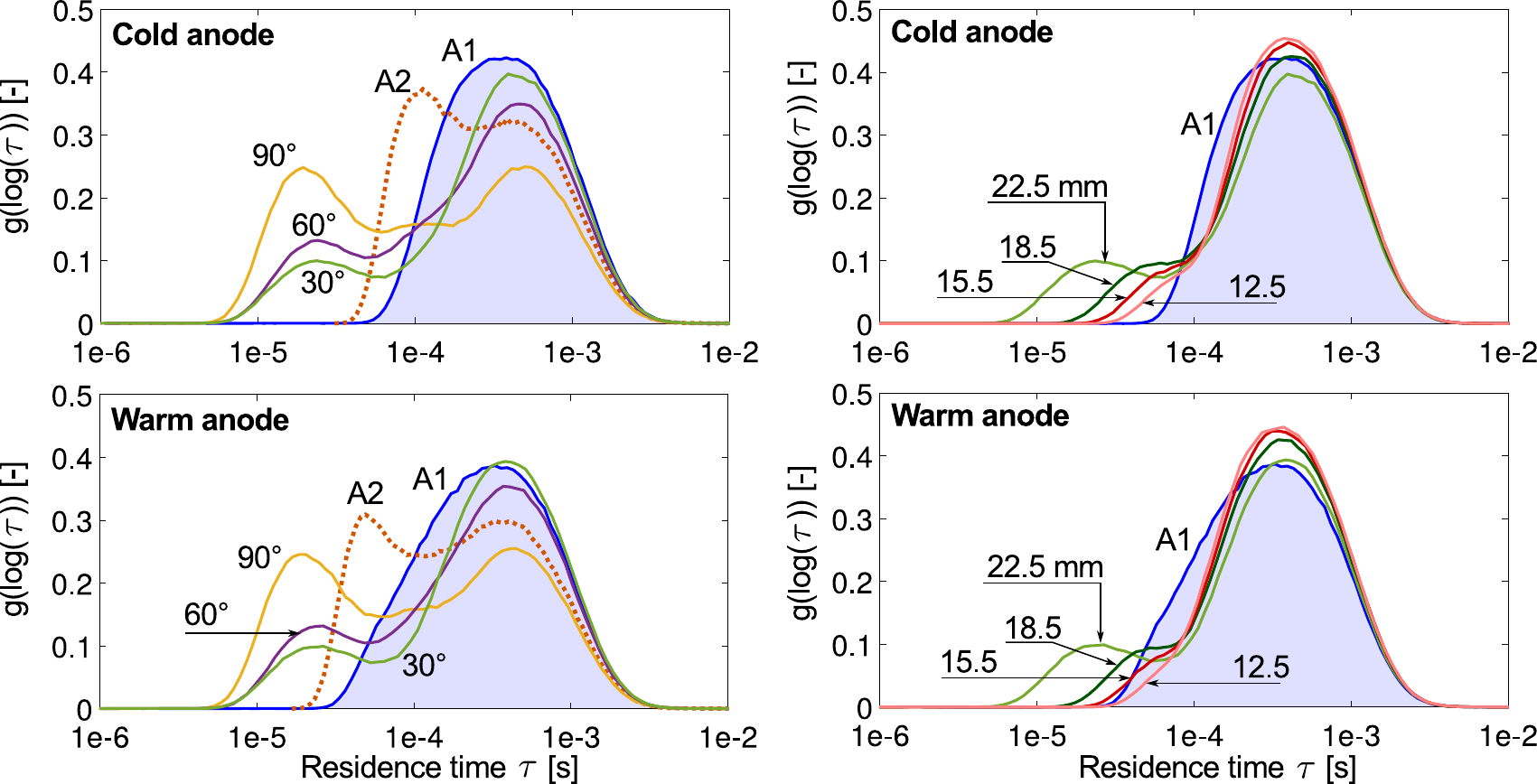}
    \caption{Normalized distribution for the logarithm of residence times in the whole channel, in the free molecular limit. 
    Effect of the injection angle (Left) and slit position (Right).}
    \label{fig:collisionless-distrib-times-wholedomain}
\end{figure*}

We start by neglecting all gas-phase collisions and consider 8 configurations: the direct injection cases A1, A2, and the reversed injection cases of Table~\ref{tab:reversed-injection-cases}.
For every particle injected in the domain, the average residence time \textit{inside the whole channel} is tracked.
The resulting distribution of residence times is shown in Fig.~\ref{fig:collisionless-distrib-times-wholedomain}.

The distribution for the uniform injection appears rather simple.
The slit injection, on the other hand, shows a larger and bi-modal shape:
the lower residence-time peak is constituted by particles that, being injected with a higher velocity at the anode slit, reach the exit without colliding with the walls.
This confirms the intuitive statement that the average velocity in the channel should be kept as low as possible for the sake of the residence time.
The very same effect is visible in the warm anode case (Fig.~\ref{fig:collisionless-distrib-times-wholedomain}-Bottom), where the high anode temperature further increases the low-residence-time bump, by increasing the amount of particles that leave the domain without wall collisions.
Indeed, considering the cold anode case, a particle injected at the (sonic) average velocity of $u_z = 177.9$ m/s takes $1.4\times10^{-4}$ s to travel a distance $L_c = 0.025$ m, while for the warm anode case, the velocity is $u_z = 324.8$ m/s, reducing the time to $7.7\times10^{-5}$ s.
This matches the position of the low-residence-times peaks for the slit injection.
Between the cold and the warm-anode cases, the high-residence-times tail is unchanged, these particles being scattered and thermally accommodated to the walls, which are set at $300$ K for both simulations.

Considering the reversed injection cases B30\_1, B60\_1 and B90\_1, one can also observe a bi-modal shape.
This time, the low-residence-time tail is constituted by the small fraction of particles with a positive axial velocity at injection due to the non-zero injection temperature.
Such particles leave the domain prematurely and are effectively wasted.
Indeed this peak: (1) decreases for more inclined injections and (2) is not affected by changes in the anode temperature.

From Fig.~\ref{fig:collisionless-distrib-times-wholedomain}-Left, it appears that all cases perform worse than the uniform injection case A1.
However, the injection at $\theta = 30^\circ$ appears the best candidate in this pool.
The performance can be increased by shifting the injection slit inwards.
The effect of different slit axial positions $z_s$ is shown in Fig.~\ref{fig:collisionless-distrib-times-wholedomain}-Right, where a fixed injection angle $\theta = 30^\circ$ is chosen.
As one could expect, moving the slit inward drastically reduces the amount of lost particles and thus increases the residence time inside the channel.
In the limit of $z_s \rightarrow 0$, one can expect to retrieve a behavior very similar to the uniform injection:
most injected particles would collide on the anode and be scattered back in the chamber.

It should be stressed that, while we considered here the geometry of the SPT-100 thruster, our collisionless results map linearly to scaled geometries such as the SPT-50 thruster.
Note that the mass flow rate does not have any influence in the present collisionless analysis, where all particles behave independently.
Nonetheless, an analysis of the Knudsen number suggests that the collisionless assumptions are not far from reality, even in the presently considered conditions (see Appendix~\ref{sec:appendix-Knudsen-num}).
The analysis performed in this section gives useful insights in the behavior of neutrals in the different configurations.
Identifying the best configuration requires further analysis and considering the finite width of the ionization region.
This is done in the next section.


\subsection{Collisional case and residence time in the ionization region}\label{sec:collisional-single-particle}

In this section, we address the full problem.
Gas-phase collisions are introduced in the test-particle Monte Carlo algorithm, and the average residence time \textit{inside the ionization region} is analyzed for a number of configurations.
The previous collisionless analysis had shown the $30^\circ$ injection cases as the best candidates, as they minimize the fraction of prematurely lost particles.
In this section we only consider such injection angle.

Differently from what we did in Section~\ref{sec:free-molec-single-particle}, the distribution of residence times will not be shown for the cases of this section.
Indeed, since we only consider the ionization region and not the whole domain, all particles leaving the channel prematurely wouldn't be accounted for, and this would make a comparison of the results much less intuitive.
Only the average residence times will be considered, to which all lost particles simply do not contribute.

First, we perform a parametric analysis of the residence time at different slit positions, for the collisionless case.
Then, we repeat the simulations for a selected number of configurations by introducing the background gas, obtained from additional DSMC simulations.
The ``residence time'' variable attached to each simulated particle is updated only when the particle is inside the ionization region $z \in [13, 18] \ \mathrm{mm}$, and does not account for the time spent by the particle either before (closer to the anode) or after it.

\begin{figure}[hbt!]
\centering
\includegraphics[width=0.5\columnwidth]{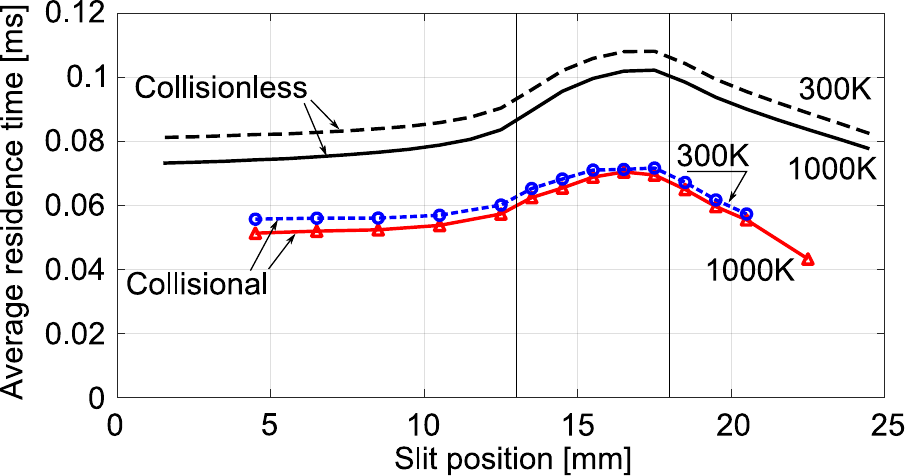}
\caption{Average residence time of neutrals in the ionization region. Vertical lines: assumed ionization region.}
\label{fig:avetime-iz-region}
\end{figure}

The average times in the ionization region are shown in Fig.~\ref{fig:avetime-iz-region}.
The effect of the anode temperature appears stronger for the collisionless cases, but much weaker when collisions are enabled. 
Considering large values of $z_s$ (slit close to the exit plane), a number of neutrals are lost prematurely, as previously observed in Section~\ref{sec:free-molec-single-particle}.
Such particles never enter the ionization region, and the residence time is correspondingly low.
On the other hand, positioning the slit between the ionization region and the anode does not bring substantial improvements.
In the limit of $z_s \rightarrow 0$, the residence time curve flattens and reaches the value obtained by direct injection.
Indeed, all particles injected too early in the channel would bounce on the anode and cross the ionization region only once, when leaving the channel.
The best improvements are obtained by centering the slit around $z_s = 16.5 \ \mathrm{mm}$, in the second half of the ionization region.
In this configuration, the maximum gain in the average residence time is $(\Delta \tau/\tau^{\mathrm{dir}})_{\%} \approx 28\%$ for the collisional case and about 40\% in the collisionless limit (representative of lower $\dot{m}_p$ and/or smaller thrusters).

\begin{figure}[htpb]
    \centering
    \includegraphics[width=0.5\columnwidth]{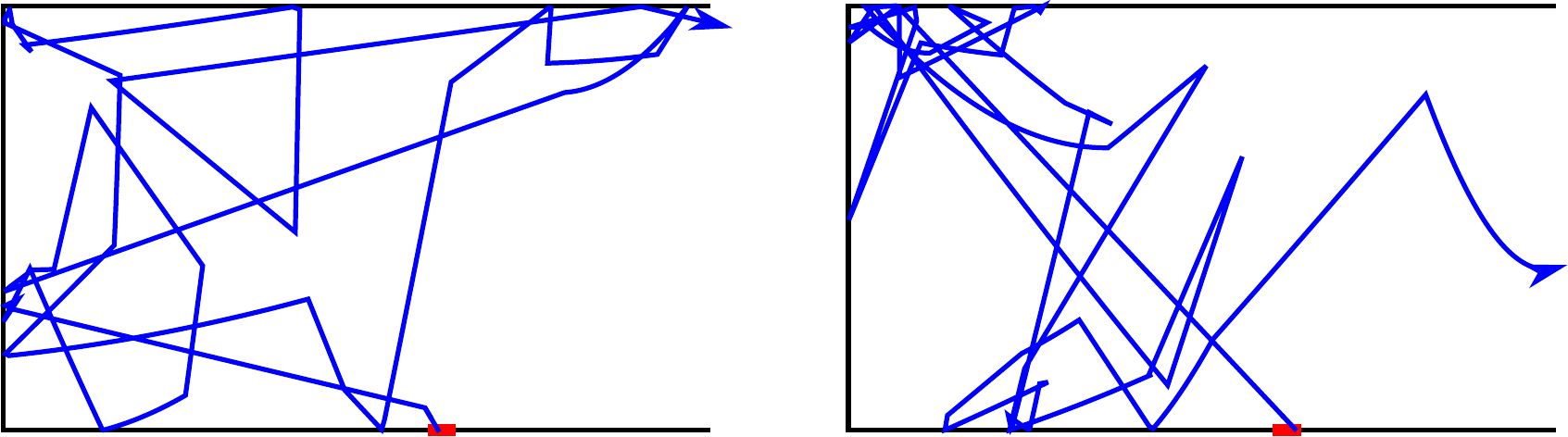}
    \caption{Dump of two sample trajectories, for case B30\_3. Gas feed is shown in red. Some trajectories appear curved due to the azimuthal motion out of the plane.}
    \label{fig:two-traj-plot}
\end{figure}

Fig.~\ref{fig:two-traj-plot} shows two sample trajectories obtained with the test-particle Monte Carlo method.
The previous considerations on the value of the Knudsen number appear evident, and one can see that the mean free path is of the same order (or slightly smaller) than the channel dimension for the considered conditions.
The free molecular regime is quickly approached for lower mass flow rates (resulting in a smaller density, and thus a larger mean free path),                
or for smaller thrusters (reducing the characteristic length); also see Appendix~\ref{sec:appendix-Knudsen-num}.

\subsection{Estimation of the mass utilization efficiency}\label{sec:estimate-efficiency}

Although electron-neutral collisions have been neglected in terms of neutral dynamics (in view of the large mass ratio), we can obtain nonetheless an estimate of the mass utilization efficiency.
This is done by assuming that, while inside the ionization region, neutral particles ionize by collisions with a cloud of background electrons with a given ionization frequency $\nu_{\mathrm{iz}}$.
Each simulated particle $i$ ionizes with a probability $P_{\mathrm{iz},i}$, that depends on the residence tau $\tau_i$ spent in the ionization region, as discussed in Eq.~\eqref{eq:Piz-1-exp-nu-tau}.
The mass utilization efficiency $\eta_m$ represents the fraction of ionized over injected particles and can thus be found from
\begin{equation}\label{eq:efficiency-avg-eq}
    \eta_m =
    \sum_{i=1}^{N_s} \frac{P_{\mathrm{iz},i}}{N_s}=
    \sum_{i=1}^{N_s} \frac{1 - \exp\left(-\nu_{\mathrm{iz}} \tau_i\right)}{N_s}\ .
\end{equation}

\begin{figure*}[htpb]
    \centering
    \includegraphics[width=1.0\columnwidth]{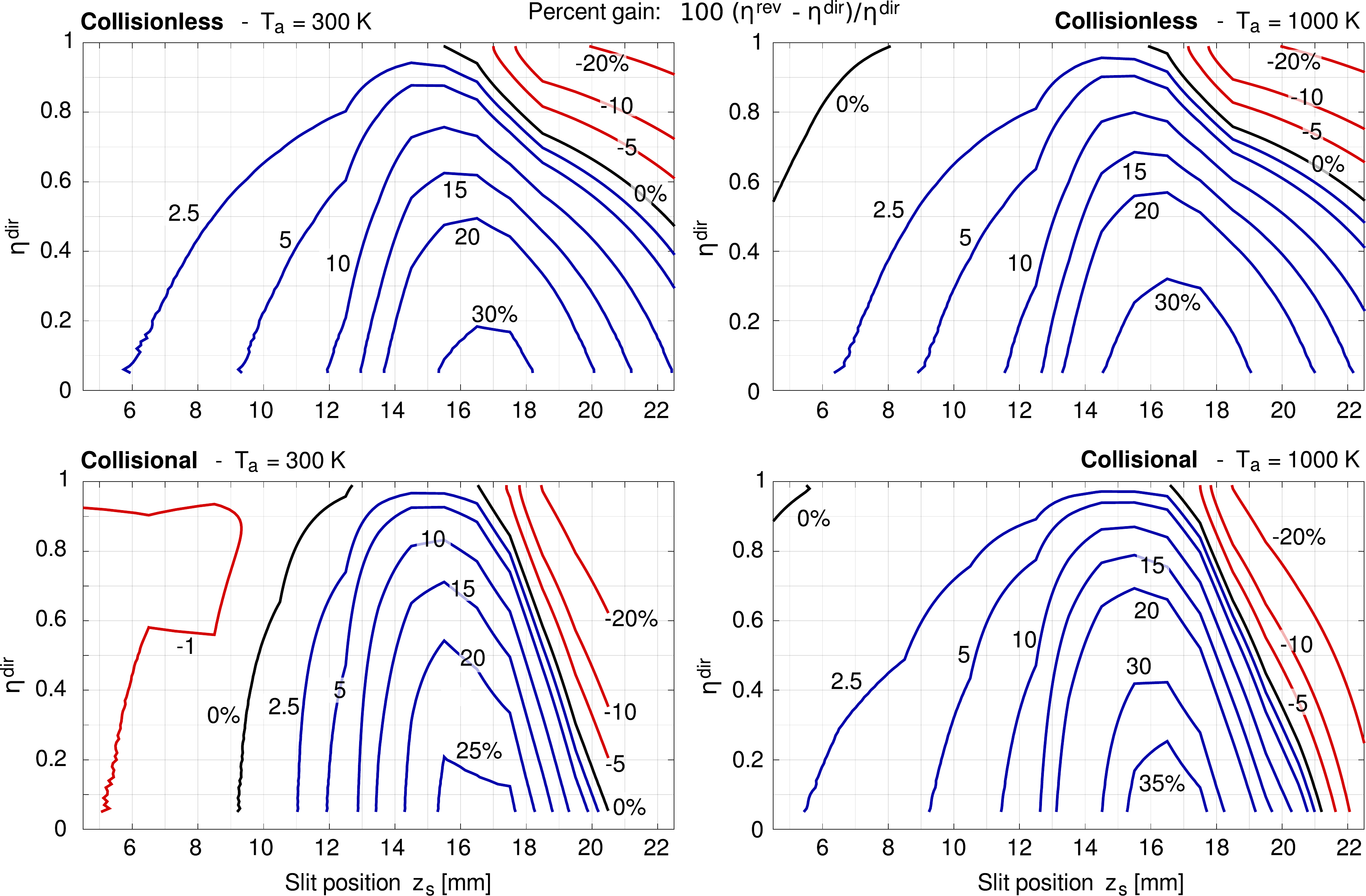}
    \caption{Percent gain for different positions of the slit center $z_s$ and different values of $\eta_{\mathrm{dir}}$. Injection angle $\theta=30^\circ$.}
    \label{fig:efficiency-plot}
\end{figure*}

Our goal is now to estimate how much improve can be obtained by use of the reversed injection strategy.
Let us first consider a thruster showing an efficiency $\eta_m^{\mathrm{dir}}$ when operated with a standard gas feed.
From a simulation of such configuration, it is possible to obtain the array of residence times $\tau_i^{\mathrm{dir}}$.
A value of $\nu_{\mathrm{iz}}$ can be then determined for such value of $\eta_m$, by solving iteratively Eq.~\eqref{eq:efficiency-avg-eq}.
The value $\nu_{\mathrm{iz}}$ just found represents an average of the the plasma state in the ionization region.
Then, a simulation of the reversed configuration is considered, and a new array of values $\tau_i^{\mathrm{rev}}$ is obtained.
By plugging such values into Eq.~\eqref{eq:efficiency-avg-eq}, together with the obtained value of $\nu_{\mathrm{iz}}$, gives the efficiency of the reversed configuration, $\eta_m^{\mathrm{rev}}$ for the same plasma conditions.
The percent gain is then computed as $100(\eta^{\mathrm{rev}}-\eta^{\mathrm{dir}})/\eta^{\mathrm{dir}}$.
Maps of the percent efficiency gain are shown in Fig.~\ref{fig:efficiency-plot} for both the collisionless and collisional cases, at different positions of the side slit.
Only the reversed injection angle of $30^\circ$ is considered.
In such maps, the collisional cases are representative of the simulated SPT-100 configuration with $\dot{m}_n=5~\mathrm{mg/s}$. 
The collisionless maps on the other hand can be directly applied to scaled versions of the SPT-100 geometry, provided that the ionization region is also in similarity.

Considering thrusters with a high initial efficiency $\eta_m^{\mathrm{dir}}>0.8$, our computations suggest that the gain to be expected lies below the 15\% in the best case scenario. 
On the other hand, for low-efficiency thrusters (say, $\eta_p\approx0.5$) the efficiency improve can reach the 20--30\%.

\section{Conclusions}\label{sec:conclusions}

In this work, we have investigated numerically the possibility to employ a reversed injection strategy as to increase the mass utilization efficiency of Hall Effect Thrusters.
In place of injecting neutrals from the anode, the injector is located along the inner wall, closer to the channel exit, and injection proceeds backwards.
Only the neutral gas flow is considered and interaction of neutral particles with electrons and ions are neglected.
Such interactions can cause an increase in neutral temperature and average velocity, but such analysis is beyond the scope of this work. 
The only plasma effect considered here is the increased anode temperature, that strongly affects the neutral dynamics.

A preliminary analysis of the reversed injection configuration is first considered, assuming heavily idealized conditions (monoenergetic particles, specular wall reflection conditions and no gas-phase collisions).
DSMC simulations were then performed assuming a more realistic diffuse wall scattering, finite temperature and enabling gas-phase collisions.
The considered geometry is that of a SPT-100 thruster, with propellant mass flow rate $\dot{m}_n = 5$ mg/s.
A baseline direct injection case with uniform injection, a slit injection and various reversed injection configurations are considered, for two different anode temperatures as to simulate more realistic operation conditions.
Results confirm a strong influence of the injection strategy on the density and average velocity fields.

In order to study more into detail the residence time of neutral particles in the channel, a test-particle Monte Carlo scheme was also employed.
From collisionless simulations, we have obtained the distributions of residence times of neutral particles inside the thruster channel.
From this analysis, a reversed injection at an angle of $30^\circ$ appears to be the best candidate among the considered configurations, as it reduces the fraction of neutrals prematurely lost in the plume.

A parametric analysis was then performed for both collisionless and collisional cases, allowing to identify the best configuration.
A slit located at $z_s=16.5~\mathrm{mm}$ appears to bring the best improvements, namely an increase of the residence time from 29\% (collisional) to 40\% (collisionless simulations) over the baseline direct injection case.
Finally, the change in mass utilization efficiency $\eta_m$ is analized for different slit locations, allowing to identify the optimal configuration for a given initial thruster efficiency.
Increases in the order of 5--15\% are to be expected for thrusters with an efficiency of $\eta_m \approx 0.8-0.9$ according to the operating conditions.
A higher gain up to 20--30\% could be expected for thrusters with a lower efficiency of $\eta_m \approx 0.5$.

As a concluding note, it should be stressed that the real performance of a gas feed configuration depends on the actual plasma properties inside the thruster, which strongly depend on the neutral flow itself.
Also, besides the beneficial increase in neutral residence time, one should carefully consider for instance the effect of a higher neutral density on electron transport \cite{book2010effect}.


\section*{Acknowledgments}

We thank Mr.~Willca Villafana (CERFACS, Toulouse), Mr.~Maciej Jakubczak (Institute of Plasma Physics and Laser Microfusion, Warsaw) and Dr.~Federico Bariselli (VKI, Sint-Genesius-Rode)
for the interesting discussions about this work.


\appendix

\section{Analysis of the Knudsen number}\label{sec:appendix-Knudsen-num}

\begin{figure*}[htpb]
    \centering
    \includegraphics[width=\textwidth]{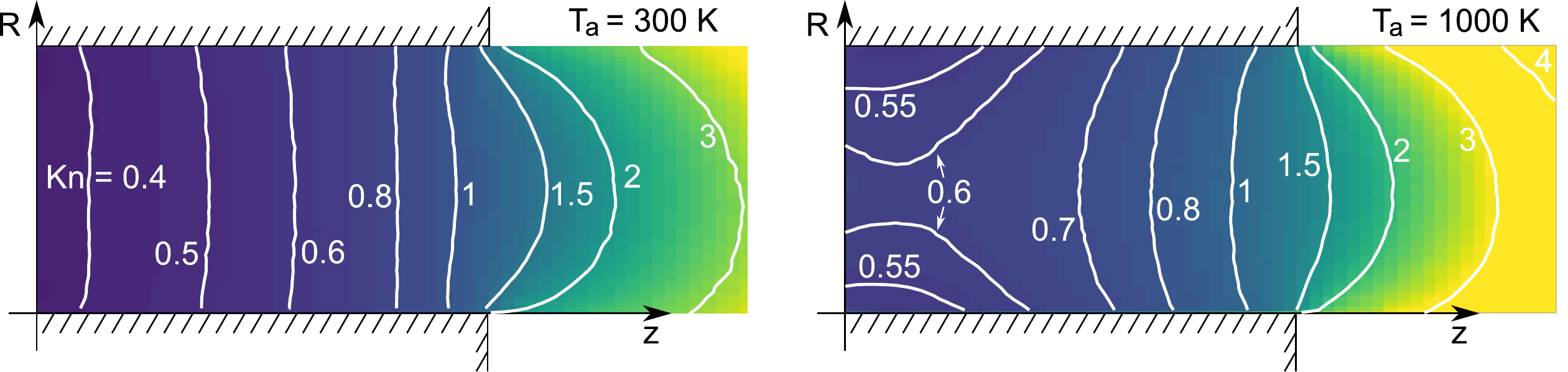}
    \caption{Contours of the local Knudsen number from the DSMC simulations of Section~\ref{sec:DSMC-simulations}, with $\dot{m}_n=5~\mathrm{mg/s}$ and for the direct uniform injection configuration. 
             Left: cold anode. Right: warm anode.}
    \label{fig:Kn-channel-uniform}
\end{figure*}

The Knudsen number $\mathrm{Kn}=\lambda/L_{\mathrm{tc}}$ gives an indication of the rarefaction regime.
Quantities $\lambda$ and $L_{\mathrm{tc}}$ are usually global values for the problem.
However, it is possible to compute maps of the local Knudsen number, by considering for example the local mean free path.
This is shown in Fig.~\ref{fig:Kn-channel-uniform} for the uniform injection case, where 
the computed Knudsen number is based on the channel width $L_{\mathrm{tc}}=0.015~\mathrm{m}$, 
and the local mean free path is computed as $\lambda=(\sqrt{2}n\sigma)^{-1}$, with $n$ the local number density
and $\sigma$ the HS cross-section for xenon.
Fig.~\ref{fig:Kn-channel-uniform}-Left shows the cold anode case, where all surfaces have the same temperature.
In such conditions, we see a uniform and continuous decrease in the local Knudsen number, that directly reflects the 
progressive gas expansion.
The anode temperature $T_a = 1000~\mathrm{K}$ is seen to increase the local Knudsen number, due to the lower associated density.
At the anode position, the warm emitted particles accommodate quickly to the colder walls, thus creating the curved contours visible in 
Fig.~\ref{fig:Kn-channel-uniform}-Right, that reflect a higher density near the walls.
It should be stressed that the present analysis of the Knudsen number is only indicative in terms of orders of magnitude.

The Knudsen number contours of Fig.~\ref{fig:Kn-channel-uniform} can be extrapolated to scaled thruster geometries and to
different mass flow rates.
In particular, for a given density, the Knudsen numbers are inverstly proportional to the thruster size, and would double by 
halving the chamber size.
Also, at the considered rarefaction conditions, the chamber density is roughly linear with the mass flow rate $\dot{m}_n$, 
for a selected injection configuration.
Hence, the mean free path $\lambda$ is proportional to $\dot{m}_n^{-1}$, and we expect the Knudsen number to 
scale linearly, as $\mathrm{Kn}\propto\dot{m}_n^{-1}$.
This linear scaling with the mass flow rate is expected to eventually break down as low Knudsen numbers are approached, in the transitional or continuum regime, but this is likely to be far from the operating conditions of Hall thrusters.

\end{document}